# Evaluation of (BH)max and magnetic anisotropy of cobalt ferrite nanoparticles synthesized in gelatin

A. C. Lima[1], M. A. Morales[1*], J. H. Araújo[1], J. M. Soares[2], D. M. A. Melo[3], A. S. Carriço[1].

[1]*Departamento de Física Teórica e Experimental, UFRN, Natal, RN 59078-970, Brazil*

[2]*Departamento de Física, UERN, Mossoró, RN 59610-210, Brazil*

[3]*Programa de Pós-Graduação em Ciência e Engenharia de Materiais, UFRN, Natal, RN 59078-970, Brazil*

*Corresponding author.

E-mail address: Marco.MoralesTorres@gmail.com (M.A. Morales)

**Abstract**

$CoFe_2O_4$ nanoparticles were synthesized using gelatin as a polymerizing agent. Structural, morphological and magnetic properties of samples treated at different temperatures were investigated by X-ray diffraction, scanning electron microscopy, Mössbauer spectroscopy and magnetization measurements. Our results revealed that the samples annealed at 623 K and temperatures above 973 K have a cation distributions given by $(Co_{0.19}Fe_{0.81})[Co_{0.81}Fe_{1.19}]O_4$ and $(Co_{0.06}Fe_{0.94})[Co_{0.94}Fe_{1.06}]O_4$, respectively. The particle sizes varied from 73 to 296 nm and the magnetocrystalline anisotropy, $K_1$, has values ranging from $2.60 \times 10^6$ to $2.71 \times 10^6$ J/m$^3$, as determined from the law of approach to saturation applied to the MxH data at high field. At 5 K, the saturation magnetization, coercive field and (BH)$_{max}$ varied from 76 to 95 Am$^2$/kg, 479.9 to 278.5 kA/m and 9.7 to 20.9 kJ/m$^3$, respectively. The reported values are in good agreement with near-stoichiometric cobalt ferrite samples.

1. **Introduction**

Recently, there is a considerable interest in research for new methods to synthesize non rare earth based hard magnetic materials. Accordingly, magnetic ferrites have drawn attention due their wide technological applications [1-2]. $CoFe_2O_4$ is a well-known magnetic material with high coercivity and remanence, moderate saturation magnetization, good chemical stability and mechanical hardness [3-4]. These characteristics make it a candidate in applications such as magnetic recording, permanent magnets, magneto-hyperthermia, magnetic drug delivery and magnetic resonance imaging [5-6]. However, to obtain Co ferrites for these applications is important to optimize their chemical composition, structure and magnetic properties. These properties are sensitive to shape and particle size as well as the occupancy of tetrahedral and octahedral sites by the metal ions [7].

Several methods to synthesize $CoFe_2O_4$ nanoparticles have been extensively studied, such as sol-gel process, thermal decomposition, co-precipitation, microemulsions, hydrothermal, and combustion reaction [8-10]. Among these, solution-phase chemical methods have attracted attention because allow to prepare nanocrystalline materials with high purity and controlled particle size at relatively low temperatures.

Recent works have reported a wet chemical method using gelatin as a polymerizing agent [11-13]. Gelatin is a protein produced by partial hydrolysis of collagen extracted from bones, connective tissues, organs of animals such as cattle, pigs, and horses. Because its solubility in water and ability to interact with metal ions in solution through the amino and carboxylic groups present in its structure, gelatin can be used as a binder gel to synthesize nanometer scale precursor particles. After the burning of this gel, nanoferrites phases can be obtained at lower temperature compared to other methods [13].

Costa et al. [11] have synthesized $CuFe_2O_4$ and $CuFeCrO_4$ nanoparticles using gelatin for application as ceramic pigments. Their results showed that the onset of crystallization of samples occurred at 625 K, but the spinel phase was obtained at temperatures above 775 K. For the $CuFe_2O_4$ and $CuFeCrO_4$ phases, the crystallite sizes ranged from 50 to 70 nm and 47 to 60 nm, respectively. Furthermore, they observed that color of pigments varied as a function of composition and heat treatment. Peres et al. [12] prepared $NiCo_2O_4$ nanoparticles using Ni and Co nitrates and commercial gelatin [12]. The results showed the formation of $NiCo_2O_4$ and $Ni_xCo_{1-x}O$ phases with sizes ranging from 20 to 100 nm, when the sample was annealed from 633 to 1223 K.

Despite several studies found in the literature, we noted that there is not any report on the synthesis of Co ferrites by chemical method using gelatin. Thus, the aim of this work is to report the preparation of Co ferrite by this method and study its chemical composition, structure and magnetic properties.

## 2. Experimental

Cobalt ferrite nanoparticles were synthesized by using $Co(NO_3)_2 \cdot 6H_2O$ (VETEC), $Fe(NO_3)_3 \cdot 9H_2O$ (VETEC) and gelatin. The masses of reagents were $6.202 \times 10^{-3}$ kg of $Co(NO_3)_2 \cdot 6H_2O$, $17.221 \times 10^{-3}$ kg of $Fe(NO_3)_3 \cdot 9H_2O$ and $5 \times 10^{-3}$ kg of gelatin. Initially, gelatin was dispersed in distilled water at 323 K. Then, $Co(NO_3)_2 \cdot 6H_2O$ and $Fe(NO_3)_3 \cdot 9H_2O$ were added to the above solution and kept under stirring at 353 K until the formation of a viscous gel. To obtain the precursor powder, the gel was heat treated at 623 K for 3h, then, the precursor was finely crushed and again annealed for 2 h at 973, 1173 and 1273 K. The samples annealed at 623, 973, 1173 and 1273 K were named S1, S2, S3 and S3, respectively.

Structural characterization of samples was carried out by X-ray diffraction (XRD) using a Mini Flex II Rigaku diffractometer and Cu Kα radiation. XRD data were collected in the 2θ range between 10° and 80° with a scan rate of 5° $min^{-1}$ and 0.02° step. Crystalline phases were identified using the ICDD database. Relative concentration of phases, lattice parameters and crystallite size were obtained by using the Rietveld refinement method. The morphology and particle size distribution of samples were analyzed on a TESCAN MIRA3 field emission scanning electron microscope (SEM). Mössbauer spectra were recorded in transmission mode, at room temperature, using a spectrometer from Wiessel with a $^{57}Co$:Rh source and activity of 25 mCi. Isomer shifts values are related to α-Fe. Magnetic measurements were performed as a function of temperature (5 to 300 K) and magnetic field (up to 10T) by using a commercial VSM - Physical Properties Measurement System (PPMS) Dynacool from Quantum Design.

## 3. Results and discussion

The XRD patterns of $CoFe_2O_4$ samples treated at different temperatures are shown in Fig. 1. Similar diffractograms were observed for all samples and were indexed to cubic spinel phase (109044-ICSD). Besides the spinel phase, a small amount of CoO (9865-ICSD) was identified as a secondary phase for sample S1. We noted that crystallite size increased with heat treatment temperature, this may happen as result of coalescence of particles [14]. The lattice parameters and crystallite sizes varied from 8.373 to 8.387 Å, and 73 to 296 nm, respectively. The lattice parameters are in good agreement with those reported in literature for cobalt ferrite[15]. Table 1 shows the results obtained from the Rietveld refinement.

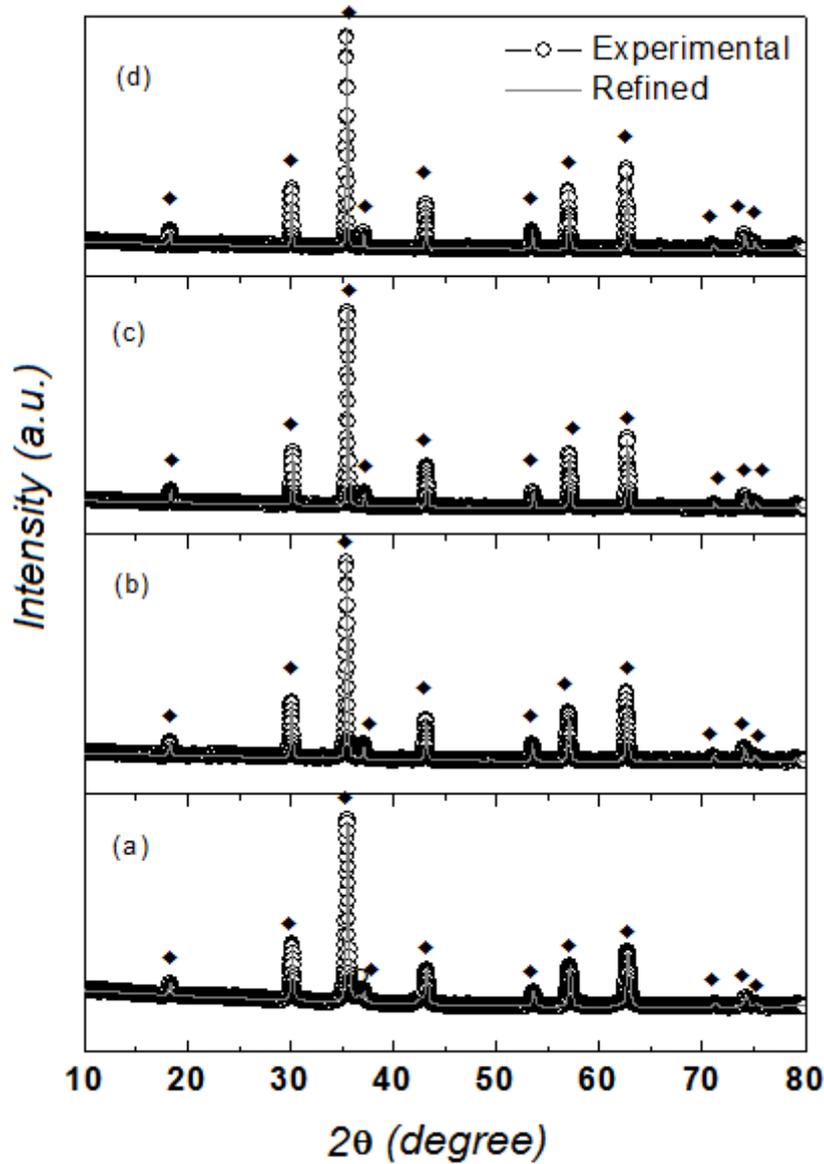

**Fig. 1.** XRD patterns of $CoFe_2O_4$ samples (a) S1, (b) S2, (c) S3 and (d) S4. Symbols are related to phases: ◆ $CoFe_2O_4$, ○ CoO

**Table 1.** Refined parameters of CoFe$_2$O$_4$ samples

| Samples | Phases % | | $D_m$ CoFe$_2$O$_4$ (nm) | Lattice parameter CoFe$_2$O$_4$ (Å) |
|---|---|---|---|---|
| | CoFe$_2$O$_4$ | CoO | | |
| **S1** | 94 | 6 | 73 | 8.373 |
| **S2** | 100 | - | 128 | 8.384 |
| **S3** | 100 | - | 189 | 8.387 |
| **S4** | 100 | - | 296 | 8.386 |

SEM images of samples S1 and S4 are shown in Fig. 2. The micrograph for sample S1 (Fig. 2a) revealed rounded particles with uniform size distribution. Elongated particles observed in figure 2b are may be due to coalescence of small particles. The sample S4 show agglomerated and pores, the porous structure is formed by escaping gases during heat treatment. In fact, gelatin provides a large amount of organic matter to the system, which may promotes the appearance of pores [11]. Inset in the top right show a histogram indicating the particle size distribution. The distributions were lognormal with average particle sizes of 53 and 308 nm. In both cases, the average particle sizes were of the same order as the ones determined from the XRD analysis.

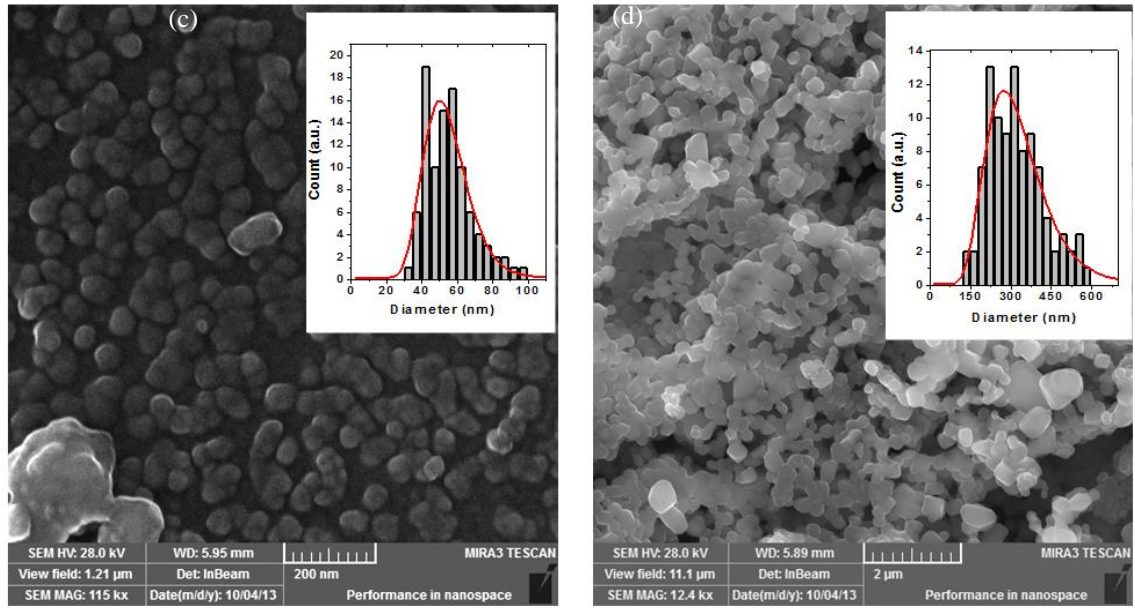

**Figure. 2**. SEM images of CoFe$_2$O$_4$ samples (a) S1 and (b) S4. Insets to the right show histograms revealing the lognormal size distribution

Mössbauer spectra (MS) recorded at room temperature are shown in Fig. 3. The MS were fitted to two sextets related to the Zeeman interaction between the hyperfine magnetic field and the nuclear magnetic moment. These subspectra are assigned to iron ions located in the tetrahedral (Fe-Tetr) and octahedral (Fe-Oct) coordination symmetry. The isomer shift (IS), hyperfine magnetic field (Hhf) and quadrople splitting (QS) values are typical for cobalt ferrite samples [16]. In these sites, $Fe^{3+}$ is coordinated by four and six oxygens. No doublet or singlet related to superparamagnetic particles or paramagnetic phases were observed. When the annealing temperature is increased, we observed a small decrease in the Fe-Oct relative absorption area (RA), accompanied by an equivalent increase in the RA of Fe-Tetr. The chemical formula unit (f.u.) of cobalt ferrites, $(Co_{1-x}Fe_x)[Co_xFe_{2-x}]O_4$, can be obtained from degree of inversion parameter (x), which is defined as the fraction of tetrahedral sites occupied by $Fe^{3+}$. In the above formula, cations enclosed in round and square brackets are ions in tetrahedral (A-sites) and octahedral (B-sites) sites, respectively. Co ferrite is completely inversed when x=1. The degree of inversion can be calculated from the ratio of subspectra areas, $RA(A)/RA(B)=f_A/f_B(x/(2-x))$, where the ratio of recoilless fraction between octahedral and tetrahedral sites at 300K is $f_B/f_A = 0.94$ [16]. Table 2 show the hyperfine parameters determined from the fits. The samples heat treated at 623 K and above 973 K, have RA(A)/RA(B) ratios of 0.73 and 0.97, respectively. The f.u. determined from these values were of $Co_{0.19}Fe_{0.81})[Co_{0.81}Fe_{1.19}]O_4$ and $(Co_{0.06}Fe_{0.94})[Co_{0.94}Fe_{1.06}]O_4$, respectively.

The linewidth attributed to the Fe-Oct is larger indicating different surroundings for Fe ions located at this site. In fact, the broadening of the B-site line was interpreted as being due to a distribution in Hhf caused by several configuration of Co and Fe nearest A-site neighbors [17]. Similar results have been observed in cobalt ferrite samples prepared by using several cooling rates [17, 18].

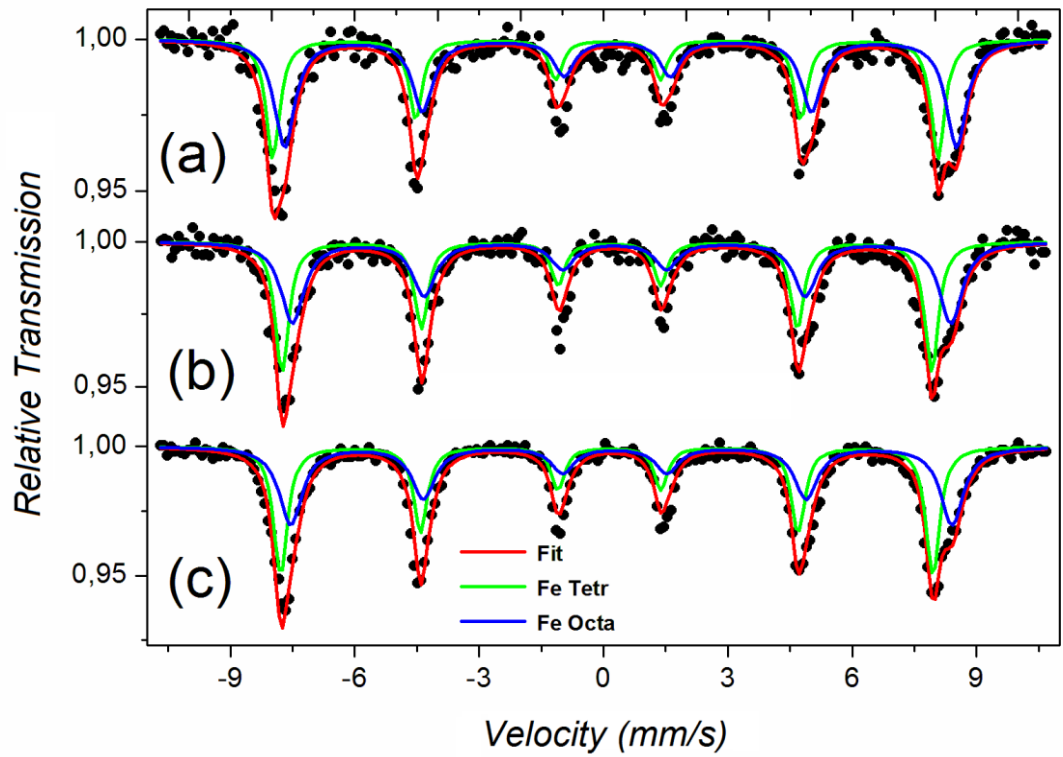

**Figure 3**. Mössbauer spectra recorded at 300 K for $CoFe_2O_4$ samples (a) S1, (b) S2 and (c) S4.

Table 2. Hyperfine parameters of $CoFe_2O_4$ samples

| Samples | Fe Sites | Hhf (T) | QS (mm/s) | IS (mm/s) | RA (%) | Linewidth (mm/s) | RA(A)/RA(B) |
|---|---|---|---|---|---|---|---|
| **S1** | Fe-Oct (B) | 50.3 | 0.18 | 0.48 | 58 | 0.60 | 0.73 |
|  | Fe-Tetr (A) | 49.8 | -0.07 | 0.18 | 42 | 0.40 |  |
| **S2** | Fe-Oct (B) | 49.3 | 0.16 | 0.46 | 51 | 0.65 | 0.97 |
|  | Fe-Tetr (A) | 48.6 | -0.06 | 0.22 | 49 | 0.40 |  |
| **S4** | Fe-Oct (B) | 49.5 | 0.16 | 0.45 | 51 | 0.67 | 0.97 |
|  | Fe-Tetr (A) | 48.8 | -0.06 | 0.22 | 49 | 0.41 |  |

Zero field cooled magnetization (ZFCM) measurements versus temperature under a field of 10 T showed a nearly constant magnetization value below 100 K, indicating that samples have reached the saturated regime. Figure 4c shows these measurements for samples S1, S2 and S4.

Magnetization versus magnetic field measurements were recorded at 5K and 300 K under an applied magnetic field of up to 10 T, figures 4a and 4b show these measurements. Conversion magnetic units from CGS to IS are 1emu/g =1Am$^2$/kg, and 1Oe = 79.58 A/m. For all the samples, the upper and lower branches of the hysteresis loops approach each other asymptotically at magnetic fields above 6 T. The anisotropy constant was determined at 5 K by fitting the high field regions (H » coercive field) to the Law of Approach to Saturation (LAS), based on the assumption that at sufficiently high field only rotational processes remain. According to the LAS, the magnetization as a function of magnetic field is usually written as [19]:

$$M = M_S \left[1 - \frac{8}{105}\left(\frac{K_1}{M_S H}\right)^2\right]$$

where the numerical coefficient 8/105 holds for random polycrystalline samples with cubic anisotropy and $K_1$ is the anisotropy constant. The anisotropy constant, $K_1$, varied from 2.60 x10$^6$ to 2.71 x10$^6$ J/m$^3$ in agreement with results obtained from magnetic torque measurements [20] and by fitting the LAS equation to MxH magnetic measurements [21] in Co ferrite samples. The high $K_1$ values are related to the strong anisotropy of Co ferrite ions and to their presence in octahedral-B sites of the spinel structure [20]. These findings were confirmed through the f.u. obtained from the MS analysis, which show that the amount of Co$^{2+}$ increases with annealing temperature. Table 3 shows the saturation magnetization values obtained from the LAS fittings. The magnetic moment (Mm) per f.u. can be determined from the Ms values. Thus, Mm=W$x$Ms$x$10$^{-3}$ /(Na$x$9.274$x$10$^{-24}$), where W is the molecular weight of Co ferrite and Na is the Avogadro's number. Therefore, the values 3.19 µ$_B$, 3.78 µ$_B$ and 3.95 µ$_B$ are the magnetic moments per f.u. for samples S1, S2 and S4, respectively. The higher magnetic moments for samples S2 and S4 reflects the higher occupancy of Co ions in the octahedral sites. These values are in agreement with the ones reported for Co ferrites with different stoichiometries and indicates the spin and orbital magnetic contribution of Co ions [17, 22].

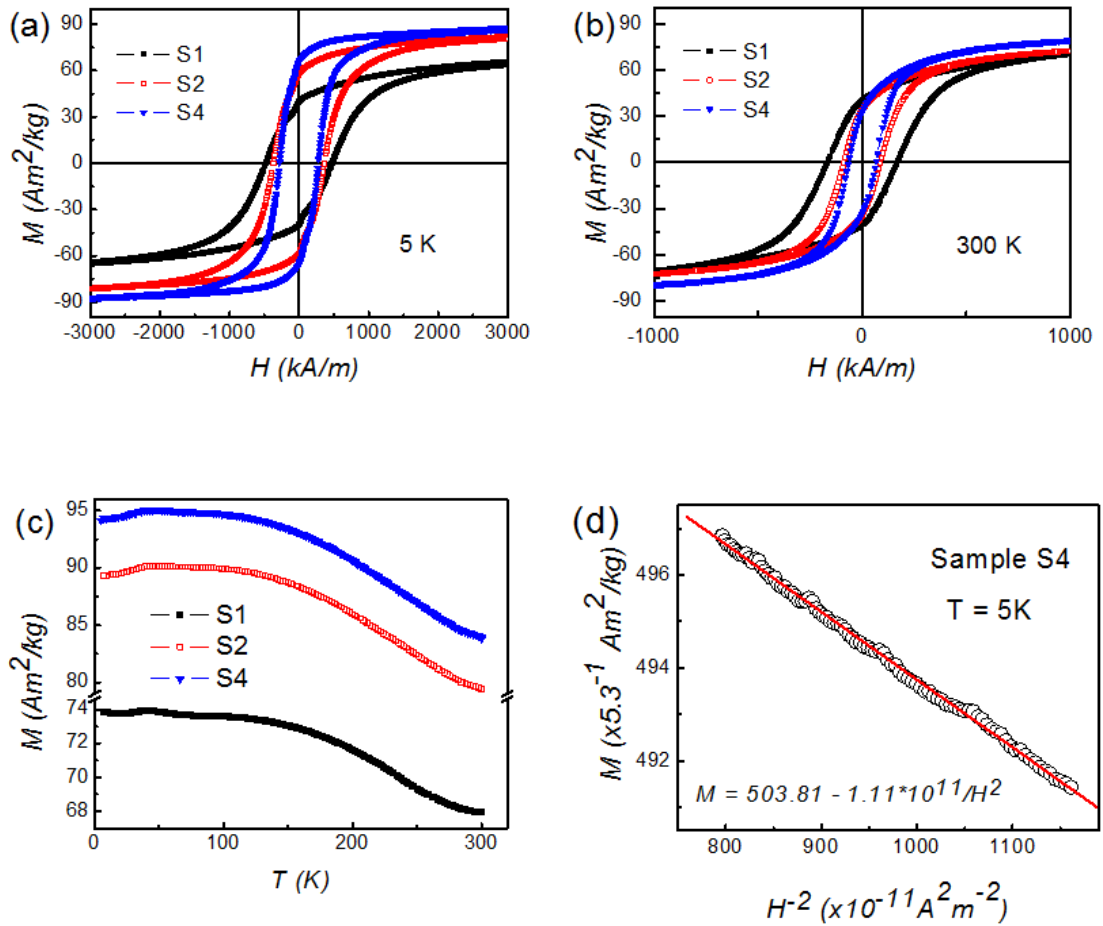

Figure 4 – Magnetization of samples S1, S2 and S4 measured at (a) 5K and (b) 300K. (c) ZFCM measurements under a magnetic field of 10 T. (d) Fit to the LAS of the high field of the MxH$^{-2}$ data for sample S4.

The magnetic parameters obtained from hysteresis loops, such as saturation magnetization ($M_s$), remanent magnetization ($M_r$), coercivity ($H_c$) and ratio $M_r/M_s$, are shown in Table 3. The values in parenthesis correspond to measurements performed at 300 K, while the others correspond to measurements performed at 5 K. The crossover from the single domain to multi domain magnetic system is related to the magnetocrystalline anisotropy. In $CoFe_2O_4$, the particle size to form a multidomain system is about 70 nm [23], crystallites exceeding this size will have reduced Hc values. Although, recent studies showed that $CoFe_2O_4$ nanoparticles with a partially inverse structure had a maximum value of Hc at size between 25-30 nm, and small values above 45 nm [24]. Our samples exhibited decreasing Hc values when particle size increased, as expected for particles with sizes above the critical diameter to form a multi domain regime.

**Table 3**. Magnetic parameters of Co ferrite samples

| Sample | $M_s$ (Am²/kg) | $M_r$ (Am²/kg) | $H_c$ (kA/m) | $M_r/M_s$ | $K_1$ (J/m³) | (BH)max (kJ/m³) |
|---|---|---|---|---|---|---|
| **S1** | 76 (72) | 40 (40) | 479.9 (167.1) | 0.53 (0.56) | $2.40 \times 10^6$ | 9.7 |
| **S2** | 90 (78) | 58 (34) | 369.3 (87.5) | 0.64 (0.45) | $2.65 \times 10^6$ | 18.8 |
| **S4** | 95 (82) | 65 (33) | 278.5 (65.3) | 0.68 (0.40) | $2.71 \times 10^6$ | 20.9 |

Parameters in parenthesis are results from measurements performed at 300 K. Other values are related to measurements recorded at 5 K.

Figure 5a shows the $B = \mu_o (H + M)$ versus H curve for sample S4 measured at 5 K. An important parameter for hard magnetic materials is the (BH)max value, which is the largest area of the rectangle that can fit in the demagnetizing B versus H curve at the second quadrant, see figure 5b. Figure 5c shows the modulus of the product B*H versus H, where the maximum value determined from this plot was 20.9 kJ/m³. The (BH)max values show a strong increase from sample S1 to S2 and from sample S2 to S4 did not change to much. This result is in agreement with the large value of remanence magnetization and high anisotropy constant for samples S2 and S4.

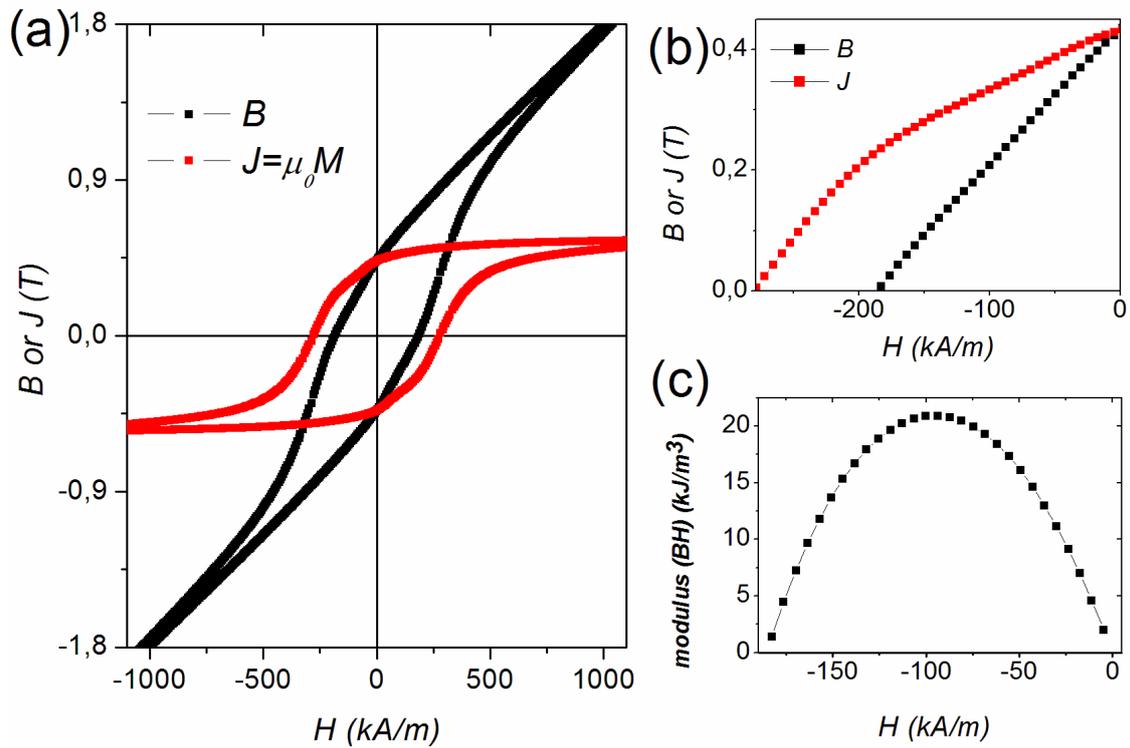

Figure 5 – Magnetization of sample S4. (a) B or J versus magnetic field H. (b) Second quadrant of the demagnetizing curve. (c) Modulus (BH) versus magnetic field.

## 4. Conclusions

The synthesis using gelatin is an alternative route, simple and inexpensive to prepare oxide nanoparticles. Cobalt ferrite nanocrystals with diameters ranging from 73 to 296 nm were prepared in gelatin. The samples have diameter compatible with multi domain magnetic particles. Mössbauer spectroscopy measurements showed that the annealing process of the cobalt ferrite nanocrystals affected the average crystallite size but also the distribution of Fe ions within A- and B-sites in the spinel structure. We found a degree of inversion close to 1.0 for samples heat treated above 973 K. The anisotropy values are in agreement with hard magnetic materials. The saturation magnetization showed magnetic moments per f.u. compatible with near-stoichiometric Co ferrite.

**5. Acknowledgments**

A.C. Lima thanks to CNPq/PNPD (561256/2010-1) by the financial support.